\documentstyle[twoside,fleqn,npb,epsfig]{article}
\def\gapp{\lower0.6ex\hbox{$\stackrel{>}{\sim}$}}

\newcommand{\AmS}{{\protect\the\textfont2
  A\kern-.1667em\lower.5ex\hbox{M}\kern-.125emS}}

\title{
\vspace{-2.5cm}                         
{\normalsize DESY 99--069}     \\[-0.2cm]      
{\normalsize HUB--EP--99/25}   \\[-0.2cm]      
{\normalsize HLRZ 1999--21}   \\[-0.2cm]      
{\normalsize June 1999}      \\              
\vspace{0.7cm} 
Higher-twist corrections to nucleon structure 
functions from
lattice QCD\,\thanks{Talk given by G. Schierholz at DIS99, April 19 -
  23, 1999, Zeuthen, Germany.}}

\author{S. Capitani\address{Deutsches Elektronen-Synchrotron DESY,
D-22603, Hamburg, Germany},
M. G\"ockeler\address{Institut f\"ur Theoretische Physik,
Universit\"at Regensburg, D-93040 Regensburg, Germany},
R. Horsley\address{Institut f\"ur Physik, Humboldt-Universit\"at 
zu Berlin, D-10115 Berlin, Germany},
D. Petters\address{Institut f\"ur Theoretische Physik, Freie
Universit\"at Berlin, D-14195 Berlin, Germany}$^{,}$%
\address{Deutsches Elektronen-Synchrotron DESY, John von Neumann-Institut
  f\"ur Computing NIC,\\\hspace{0.155cm}D-15735 Zeuthen, 
Germany}, D. Pleiter$^{\rm d,e}$, P. Rakow$^{\rm b}$ and G.
Schierholz$^{\rm a,e}$}

\begin{document}

\begin{abstract}
A genuinely non-perturbative evaluation of higher-twist contributions
to the structure functions of the nucleon, with all mixing effects and
renormalon ambiguities taken care of, is presented. Higher-twist
corrections turn out to be significant at moderate values of $q^2$.
\vspace{-0.3cm}
\end{abstract}

\maketitle

\section{INTRODUCTION}

The calculation of power corrections to the deep-inelastic structure
functions of the nucleon is a long-standing problem. Recent
phenomenological studies~\cite{Liuti} have shown that they account
for more than 10\% of the lower moments of $F_2$ at $q^2 = 5 \;
\mbox{GeV}^2$. For the theoretical understanding of the structure
functions, in particular their evolution as a function of $q^2$, it is
thus important to gain quantitative control of these effects. 

The theoretical basis for the calculation is the operator product 
expansion (OPE):
\begin{eqnarray}
W_{\mu\nu} \hspace{-0.2cm} &\equiv& \hspace{-0.2cm} 
           \langle p|J_\mu(q)J_\nu(-q)|p \rangle \nonumber \\[-0.6em]
           & & \label{ope} \\
           \hspace{-0.2cm} &=& \hspace{-0.2cm} 
           \sum_{m,n} C_{\mu\nu,\mu_1\cdots\mu_n}^m(\mu^{-1})\, 
               \langle p|{\cal O}_{\mu_1\cdots\mu_n}^m(\mu^{-1})|p \rangle ,
            \nonumber 
\end{eqnarray}
where $m$ distiguishes different operators with the same Lorentz symmetries.
The Wilson coefficients $C$ are universal and depend only on $q$ and the
renormalization point $\mu$, while all dependence on the target momenta is  
contained in the operator matrix elements. The product of Wilson coefficients
and matrix elements does not depend on $\mu$, nor
on the renormalization scheme.

Besides the familiar operators of twist two, such as 
\begin{displaymath} 
{\cal O}_{\mu_1 \cdots \mu_n} \!= \bar{\psi}\gamma_{\{\mu_1}D_{\mu_2}\cdots
D_{\mu_n\}}\psi - \mbox{traces},
\end{displaymath}
the OPE (\ref{ope}) receives contributions from twist-four and higher
operators, which give rise to power corrections, and which so far have been
neglected 
in the theoretical analysis. A typical twist-four operator is 
\begin{displaymath} 
{\cal O}_{\mu_1 \cdots \mu_n} \!= \bar{\psi}\gamma_{\mu_1}D_{\mu_2}\cdots
D_{\lambda}D_{\lambda}\cdots D_{\mu_n}\psi . 
\end{displaymath}

In addition to the calculation of matrix elements of higher-twist operators,
the evaluation of power corrections requires a non-perturbative computation
of the Wilson coefficients. This is necessary because of renormalon
ambiguities and 
mixing effects, which tie Wilson coefficients and higher-twist matrix elements
together~\cite{renorm}.   

In \cite{qcdsf} we have developed a method that allows a lattice calculation
of the Wilson coefficients. With this method we can compute the r.h.s. of
(\ref{ope}) entirely on the lattice: 
\begin{displaymath}
\sum_{m,n} C_{\mu\nu,\mu_1\cdots\mu_n}^m(a)\, 
\langle p|{\cal O}_{\mu_1\cdots\mu_n}^m(a)|p \rangle ,
\end{displaymath}
$a = \mu^{-1}$ being the lattice cut-off, without having to introduce an
intermediate (renormalization) scale parameter. This approach automatically 
takes care of all mixing effects, and it avoids renormalon ambiguities
altogether.

In this talk we shall present first results of an all-lattice calculation of
higher-twist corrections to the structure function of the nucleon.

\section{WILSON COEFFICIENTS}

Let us concentrate on the Wilson coefficients first. The basic
idea~\cite{qcdsf} is to
compute $W_{\mu\nu}$, and the operator matrix elements in the OPE
going with it, for off-shell quark states in Landau gauge, and to extract the
Wilson coefficients from this information. Because the quarks are off-shell
(and euclidean), we are far from the Bjorken limit, so that higher-twist
operators are not suppressed.  

\begin{figure}[t]
\vspace{-0.9cm}
\hspace*{-0.3cm} \epsfig{figure=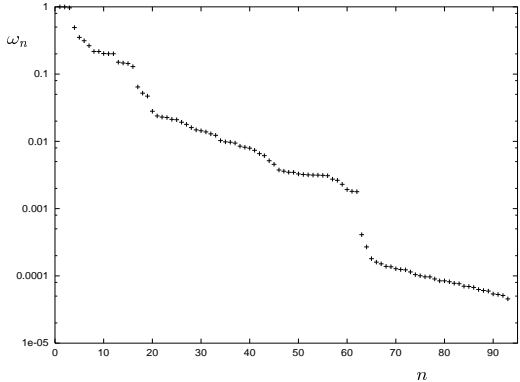,height=6.5cm,width=8.5cm}
\vspace{-1.7cm}
\caption{The eigenvalues $\omega_n$ against $n$.}
\label{fig1}
\vspace{-0.5cm}
\end{figure}

We consider a system of quark states with momenta 
$p_m$, $m = 1, \cdots, M$. 
The photon momentum $q$ is held fixed. We drop the Lorentz indices,
and we label the operators and the associated Wilson coefficients
by $n = 1, \cdots, N$. The
matrix elements of the operators between states of
momentum $p_m$ are denoted by $O_n^{p_m}$, and the corresponding current
matrix elements are called $W^{p_m}$. Note that both, $O_n^{p_m}$ and $W^{p_m}$
are $4\times 4$ matrices.
The problem is then to solve the $N \times (M \times 16)$ system
of equations   
\begin{equation}
\left(
\begin{array}{ccc}
 O_{1}^{p_1}  &  \hspace*{-0.27cm}\cdots\hspace{-0.27cm} & 
 O_{N}^{p_1}   \\
\vdots       & \hspace*{-0.27cm} \hspace{-0.27cm} & 
\vdots   \\
 O_{1}^{p_M}  & \hspace*{-0.27cm}\cdots\hspace*{-0.27cm} & 
 O_{N}^{p_M}   \\
\end{array}
\right)
\left(
\begin{array}{ccc}
C_{1}  \\
\vdots \\
C_{N}  \\
\end{array}
\right)
=
\left(
\begin{array}{ccc}
W^{p_1}  \\
\vdots    \\
W^{p_M}  \\
\end{array}
\right)
\label{matrix}
\end{equation}
for the Wilson coefficients $C_n$. In the following we take into account
all quark-bilinear operators with up to three covariant derivatives.

\begin{figure}[t]
\vspace{-0.9cm}
\hspace*{-0.3cm} \epsfig{figure=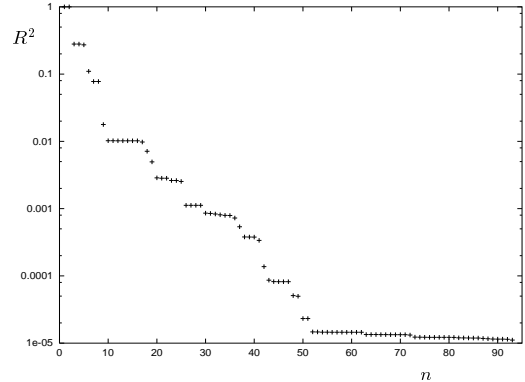,height=6.5cm,width=8.5cm}
\vspace{-1.7cm}
\caption{The residual error $R^2$ against $n$.}
\label{fig2}
\vspace{-0.5cm}
\end{figure}

Generally, not all operators are independent, so that the system is 
overdetermined. Let us write (\ref{matrix}) as 
\begin{equation}
O\, C = W.
\label{ocw}
\end{equation}
To solve (\ref{ocw}), we apply a singular value
decomposition, which is the standard method for solving overdetermined sets of
equations~\cite{nr}. Writing 
\begin{displaymath}
O= U\,\omega \,V^T,
\end{displaymath}
where $U$ ($V$) is a column-orthonormal $(16 \times M) \times N$
($N\times N$) matrix, and 
$\omega = \mbox{diag}(\omega_n)$ with positive real eigenvalues $\omega_n$
arranged in descending order, the solution to (\ref{ocw}) is  
\begin{displaymath}
C= V\,\mbox{diag}(1/\omega_n)\,U^T \: W.
\end{displaymath}
In a calculation with infinite precision one would find $\omega_n = 0$ for 
any $n$ exceeding the number of independent operators, and in these instances
one would replace $1/\omega_n$ by zero. In a calculation like ours we can at
most hope for a sharp drop of the $\omega$'s down to the level of the noise.

In total there are 1360 operators to be considered, which by symmetry
reduce to 93. We have considered 70 different momenta, thus resulting in a 
$93\times 1120$ system of equations. The calculations are
done on a $32^4$ lattice at $\beta = 6$ using Wilson fermions.  
In Fig.~1 we show a typical picture of the eigenvalues
$\omega_n$. We see a sharp drop at $n \approx 63$. In Fig.~2 we show the
residual error $R^2 = |W - OC|^2/|W|^2$ as a function of $n$ with 
$1/\omega_{n+1},\cdots, 1/\omega_N$ set to zero. We see that the error does not
decrease significantly anymore for $n \,\gapp\, 63$, indicating that we have
`hit the noise', and that we may truncate the system at this value of $n$, what
we will do.

\section{STRUCTURE FUNCTION RESULTS}

The next step is to compute the nucleon matrix elements of the
operators entering the OPE (\ref{ope}). This, combined with the Wilson
coefficients 
evaluated in the last section, then gives us the desired hadronic
tensor $W_{\mu\nu}$ of the nucleon. For this calculation it is
sufficient to consider a few different nucleon momenta only.

From $W_{\mu\nu}$ we can derive moments of the nucleon structure
functions. Here we shall restrict ourselves to the lowest
non-trivial moment of the unpolarized structure function. Projection
onto spin two states gives~\cite{otto}
\begin{equation}
\int\!\! {\rm d}\theta \sin^2\!\theta\, C_2^1(\cos\theta)
W_{\lambda\lambda}\! \propto\!\! \int\!\!{\rm d}x \,(2xF_1\! - \! F_L/2),
\label{mom}
\end{equation} 
where $C_2^1(\cos\theta)$ is a Gegenbauer polynomial, and
$\cos\theta = pq/|p||q|$. Similar results are 
obtained for other combinations of $\mu, \nu$.

The nucleon matrix elements are computed on a smaller, $16^3 32$
lattice (at the quoted value of $\beta$), and the combined results are
extrapolated to the 
chiral limit. The calculations are done in the quenched approximation,
so that our numbers refer to non-singlet structure functions.

We denote the r.h.s. of (\ref{mom}) by ${\cal M}_2(q^2)$. In the
language of the parton model
this corresponds to the moment $\langle x \rangle$. We write 
\begin{displaymath}
{\cal M}_2(q^2) = m_2 + c_2/q^2,
\end{displaymath}
separating the power correction from the leading, logarithmically
varying contribution. This makes sense, because we can identify those 
higher-twist contributions which are logarithmically behaved, and those which
are power behaved.

We present our results in Fig.~3. All our numbers should be regarded
as preliminary. The first observation we make is that power
corrections are large. They are positive, and at $q^2 = 5 \,
\mbox{GeV}^2$ they amount to 
$\approx 30\%$ of the total contribution. The second, and equally
important observation is that there is strong mixing between operators of
twist two and four. This effect reduces the leading contribution ($m_2$)
substantially, if compared with the result of our previous
calculation~\cite{qcdsf2} based on perturbative Wilson coefficients and
renormalization constants. We now find much better agreement of this
quantity with
the phenomenological valence quark distribution functions.

\begin{figure}[t]
\vspace{-0.2cm}
\hspace*{0.3cm} \epsfig{figure=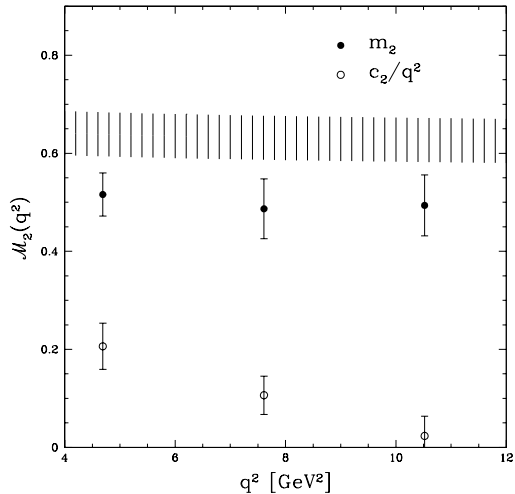,height=7cm,width=7cm}
\vspace{-0.7cm}
\caption{${\cal M}_2(q^2)$ for the proton as a function of $q^2$. The
  hatched strip shows our old result~\cite{qcdsf2} (where the width
  of the strip indicates the error) without
  higher-twist effects included.}
\label{fig3}
\vspace{-0.6cm}
\end{figure}

\end{document}